\documentclass{appolb}
\usepackage{graphicx}

\usepackage{float}
\usepackage{subfig}
\usepackage{authblk}
\usepackage{amsmath,bm}

\newcommand{\rf}[1]{(\ref{#1})}
\newcommand{\beq}{\begin{equation}}
\newcommand{\beql}[1]{\beq\label{#1}}
\newcommand{\eeq}{\end{equation}}
\newcommand{\bea}{\begin{eqnarray}}
\newcommand{\eea}{\end{eqnarray}}

\newcommand{\cD}{{\cal D}}

\newcommand{\cM}{{\cal M}}

\newcommand{\cT}{{\cal T}}

\newcommand{\cZ}{{\cal Z}}

\begin{document}
\title{Quantum gravity on a torus%
\thanks{Presented at the 6th conference of the Polish Society on Relativity}%
}
\author{Jakub Gizbert-Studnicki
\address{The M. Smoluchowski Institute of Physics, Jagiellonian University
\newline
 ul. prof. Stanis\l awa \L ojasiewicza 11, Krak\'ow, PL 30-348
 \newline
 Email: jakub.gizbert-studnicki@uj.edu.pl}
}
\maketitle
\begin{abstract}
Causal Dynamical Triangulations (CDT) is a non-perturbative lattice approach to quantum gravity where one assumes space-time foliation into spatial hyper-surfaces  of fixed topology. Most of the CDT results were obtained for  the spatial topology of the 3-sphere. It was shown that CDT has   rich phase structure, including the semiclassical phase consistent with Einstein's general relativity. Some of the  phase transitions were found to be second (or higher) order which makes a possibility of taking continuum limit viable.    Here we present new results of changing the spatial topology to that of the 3-torus. We argue that the topology change does not change the phase structure nor the order of the phase transitions. Therefore CDT results seem to be universal independent of the topology chosen.
\end{abstract}
  
\section{Introduction}

Among many possible candidate theories of {\it quantum gravity}, Causal Dynamical Triangulations (CDT) is a background independent, non-perturbative  approach based on  the Feynman's path integral formalism. CDT defines a (formal)  path integral of quantum gravity (QG)
\beql{ZCDT}
\cZ_{QG}= \int \cD_{\cM}{[g]} e^{iS_{HE[g]}}\quad \rightarrow \quad \cZ_{CDT} = \sum_{\cT}e^{iS_R[\cT]} \ ,
\eeq
by a sum over  triangulations $\cT$ constructed from four-dimensional simplicial building blocks. In CDT one additionally requires that each triangulation  $\cT$ admits a global proper time foliation into spatial  hyper-surfaces  (slices) of fixed three-dimensional topology.  Such triangulations can be constructed by using  two types of 4-simplices, called the $(4,1)$ and  the $(3,2)$ simplex.\footnote{Each 4-simplex has exactly 5 vertices. Due to the imposed proper time foliation each vertex in the triagulation has a uniquely defined (integer) time coordinate $t$ and the numbers $(n,m)$ denote the number of vertices  in $t$ and $t\pm 1$, respectively.}
In equation~\rf{ZCDT} $S_R$ is the discretized Hilbert-Einstein action $S_{HE}$ obtained  following Regge's method for describing piecewise linear geometries \cite{Regge:1961px}
\beql{SRegge}
S_{R}[\cT]=-\left(\kappa_{0}+6\Delta\right)N_{0}+\kappa_{4}\left(N_{(4,1)}+N_{(3,2)}\right)+\Delta \ N_{(4,1)} \  ,
\eeq 
where $ N_{(4,1)}$,  $ N_{(3,2)}$ and $N_0$ denote respectively the number of $(4,1)$-simplices, $(3,2)$-simplices and vertices in a triangulation $\cT$.
$\kappa_{0}$, $\Delta$ and $\kappa_4$ are three  dimensionless bare coupling constants related to the Newton's constant, the cosmological constant and the  asymmetry between lengths of space-like and time-like links in the triangulation. 
The existence of the global foliation means that each triangulation can be analytically continued between the Lorentzian and the Euclidean geometry. In the Euclidean formulation  memory of the time direction and foliation is preserved. Wick rotation changes the Lorentzian action into the Euclidean action and the path integral $\cZ_{CDT} $  becomes a partition function of (triangulated)  random geometries which can be studied numerically using Monte Carlo methods. 

\begin{figure}[H]
\centerline{
\includegraphics[width=8.5cm]{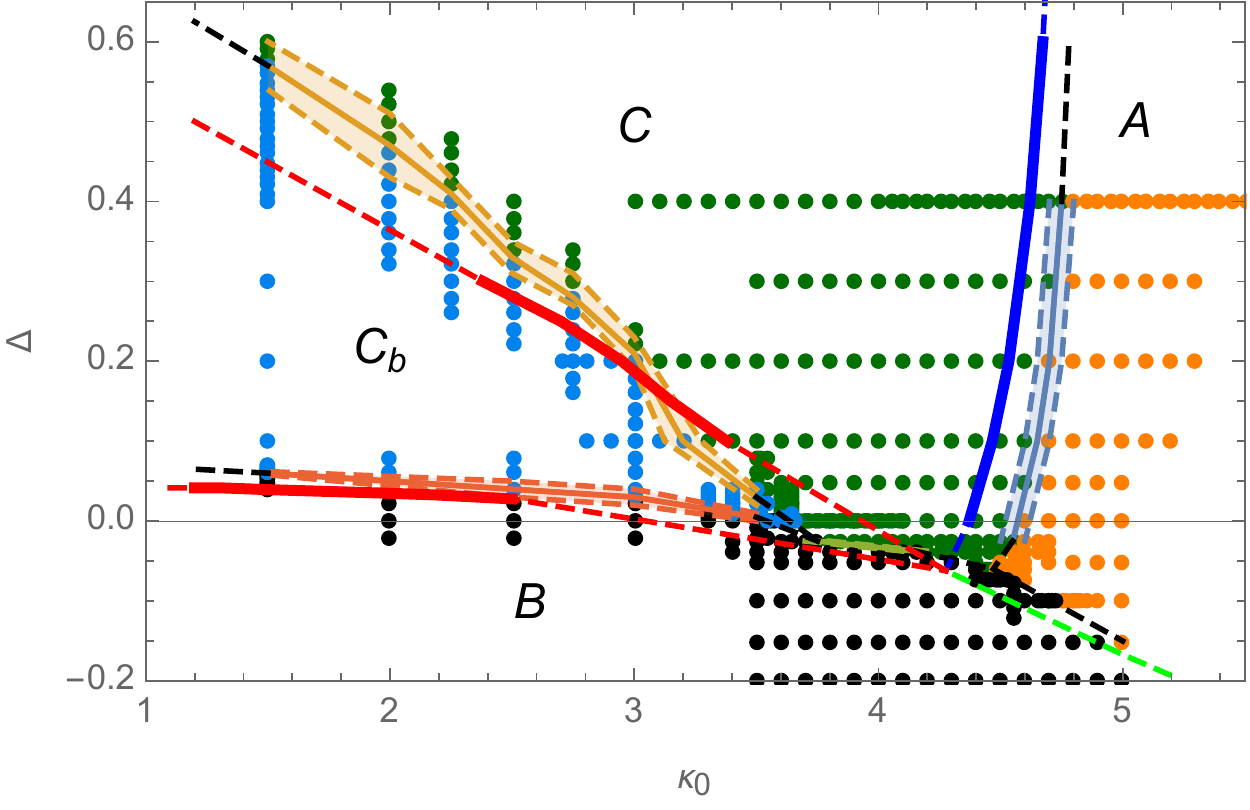}}
\caption{Phase structure of 4-dim CDT with the toroidal topology of spatial slices in the $(\kappa_0,\Delta)$ bare couplings plane. Points are actual  measurements and thin solid lines represent the measured phase transitions   (shaded regions are error bars). For comparison we also plot  the phase transitions measured in the spherical spatial topology (thick  solid lines).}
\label{Fig:F1H}
\end{figure}

Most of the previous results were obtained for  the fixed spatial topology of the 3-sphere and the studied systems were also assumed to be periodic in the (Euclidean) time. In such a case CDT showed rich phase structure (see figure \ref{Fig:F1H}) including the {\it semiclassical} phase $C$, where one observes dynamical emergence of 4-dimensional background geometry consistent with general relativity \cite{Ambjorn:2007jv} and quantum fluctuations of spatial volume are described by Hartle-Hawking minisuperspace action \cite{Ambjorn:2008wc, Ambjorn:2011ph}. Another important result was the existence of phase transitions of higher order  \cite{Ambjorn:2012ij, NewPhase} which makes a possibility of taking continuum limit viable \cite{Ambjorn:2014gsa, Ambjorn:2016cpa}.

\section{CDT with toroidal spatial topology}

Below we present new results obtained for the fixed spatial  topology of the 3-torus (and the time periodic boundary conditions), which has been studied recently. In \cite{Ambjorn:2016fbd, Ambjorn:2017} it was shown that in such a case one can observe the analogue of  phase $C$, albeit the quantum fluctuations occur around a different semiclassical background geometry than in the spherical topology.\footnote{The average spatial volume profile   in phase $C$ changes from that of the (Euclidean) de Sitter space in the spherical CDT to the flat profile in the toroidal CDT.} In the toroidal case the  quantum fluctuations of spatial volume are again well described by the minisuperspace action, and one can also observe a  quantum correction in the effective potential, which could not be  measured in the spherical CDT. 

By analyzing the behaviour of four order parameters in various points on the toroidal CDT phase diagram one can as well observe the analogues of all other phases discovered in the spherical CDT \cite{Ambjorn:2018}. One can also measure precisely the position and order of the phase transitions. The phase diagram in figure   \ref{Fig:F1H} shows that  the toroidal CDT phase transitions are only slightly shifted versus the spherical CDT case which most likely results from different finite size effects  in these topologies.\footnote{The minimal possible triangulation of the 3-torus is much larger than the minimal triangulation of the 3-sphere \cite{Ambjorn:2016fbd}.} As an additional bonus, in the toroidal case one was able to perform precise numerical simulations in the most interesting region of the parameter space where the phases meet, which was not possible in the spherical CDT. The measurements  showed that instead of the conjectured "quadruple" point, where all four phases were supposed to meet, one observes two separate "triple" points, where three phases meet. In between the two "triple" points one can also measure the direct $C-B$ phase transition, which was shown to be first order \cite{Ambjorn:2019a}, albeit with some untypical properties suggesting that the end points can be higher order.

The recent studies of the toroidal CDT confirmed that the $A-C$ transition is first order  \cite{Ambjorn:2019b} and that the $B-C_b$ transition is second (or higher) order, exactly as it was observed  in the spherical topology. For the $C-C_b$ transition, which was shown to be second (or higher) order in the spherical case, in the toroidal  CDT one observes strong hysteresis,  suggesting that the order of the transition might have changed.  So far, due to the hysteresis, the numerical algorithms used do not allow us to  make precise finite size scaling analysis of that transition so one can neither prove or disprove this hypothesis.  The order of  measured phase transitions  is summarized in the table below.

\begin{table}[H]
\begin{center}
\begin{tabular} {|c|c|c|c|}
\hline
{Phase transition}	& {Topology: $S^1\times T^3$ }	& {Topology: $S^1\times S^3$ }	 \\ \hline
\hline
$A-C$ &  1st order	& 1st order \\  \hline
$B-C$ &  1st order	&?  \\  \hline
$B-C_b$ &  higher order	& higher order \\  \hline
$C-C_b$ &  ?	& higher order \\  \hline
\end{tabular}
\end{center}
\end{table}

\section{Summary and conclusions}
In principle the CDT results may depend on the choice of (fixed) spatial topology.  Most of the previous studies of CDT were done for the spatial topology of the 3-sphere and the time periodic boundary conditions. We have briefly presented the recent results of CDT with the spatial topology of the 3-torus. We have shown that the phase structure and the order of the measured phase transitions have not changed due to the topology change. The studies provide evidence that the CDT results are universal independent of the real topology of the Universe.

\section{Acknowledgements}
The author  acknowledges  support of the grant UMO-2016/23/ST2/00289 from the National Science Centre Poland. The results described herein were obtained in  collaboration with J. Jurkiewicz, J. Ambj\o rn,  A. G\"orlich, D.N.~Coumbe, Z.~Drogosz,  D. N\'emeth, and G. Czelusta.

\end{document}